\documentstyle[12pt]{article}
\newcounter{mycount}

\newcommand{\f}{\frac}
\newcommand{\half}{\frac{1}{2}}

\newcommand{\p}{\partial}

\newcommand{\A}{$SH_N (\nu )\ $}
\newcommand{\be}{\begin{equation}}
\newcommand{\ee}{\end{equation}}
\newcommand{\bee}{\begin{eqnarray}}
\newcommand{\eee}{\end{eqnarray}}
\newcommand\nn{\nonumber \\}
\newcommand\n{ }
\newcommand\noi{\noindent}
\newcommand{\ie}{$i.e.\ $}
\newcommand{\eg}{$e.g.\ $}

\newcommand{\ig}{\Big}
\newcommand{\igg}{\Bigg} 
\font\frtnfr=eufm10   scaled\magstep1
\font\twlfr=eufm10
\font\tenfr=eufm10

\newfam\frfam
\textfont\frfam=\frtnfr
\scriptfont\frfam=\twlfr
\scriptscriptfont\frfam=\tenfr
\def\fr{\fam\frfam}

\font\frtnopen=msbm10  scaled\magstep2
\font\twlopen=msbm10
\font\tenopen=msbm10

\newfam\openfam
\textfont\openfam=\frtnopen
\scriptfont\openfam=\twlopen
\scriptscriptfont\openfam=\tenopen
\def\open{\fam\openfam}

\font\frtnsf = cmss12 scaled\magstep1
\font\twlsf = cmss10
\font\tensf = cmss9

\newfam\Scfam
\textfont\Scfam = \frtnsf
\scriptfont\Scfam = \twlsf
\scriptscriptfont\Scfam = \tensf

\begin{document}
\renewcommand{\theequation}{\arabic{equation}}
\bibliographystyle{nphys}
\setcounter{equation}{0}
\phantom{qq}\ \\
\vskip 20 mm
\centerline{\Large\bf Supertraces on the Algebras of Observables of the}
\centerline{\Large\bf Rational Calogero Model with Harmonic Potential}
\vskip 20 mm

\centerline{S.E.~Konstein and M.A.~Vasiliev}

\vskip 5 mm
\noi
\centerline{I.E.Tamm Department of Theoretical Physics,
P. N. Lebedev Physical Institute,} \\
\centerline {117924
Leninsky Prospect 53, Moscow, Russia.}
\vskip 3mm \noi
\vskip 19mm
\renewcommand{\theequation}{\arabic{equation}}
\setcounter{equation}{0}
\begin{abstract}
We define a complete set of supertraces on the algebra $SH_N(\nu )$, the
algebra of observables of the $N$-body rational Calogero model with harmonic
interaction. This result extends the previously known results for the
simplest cases of $N=1$ and $N=2$ to arbitrary $N$. It is shown that \A admits
$q(N)$ independent supertraces where $q(N)$ is a number of partitions of $N$
into a sum of odd positive integers, so that $q(N)>1$ for $N\ge 3$. Some
consequences of the existence of several independent supertraces of $SH_N
(\nu )$ are discussed such as the existence of ideals in associated
$W_{\infty}$ - type Lie superalgebras.  \end{abstract}

\section{Introduction}

In this paper we investigate some properties of the associative algebras
which were shown in \cite{1,2,3} to underly the rational Calogero model
\cite{4} and were denoted as \A in \cite{5}. Algebra $SH_N (\nu )$ is the
associative algebra of polynomials constructed from arbitrary elements
$\sigma$ of the symmetric group $S_N$ and the generating elements
$a^\alpha_i\,$ obeying the following relations \be \label{begin}
\sigma\,a^\alpha_i=a^\alpha_{\sigma(i)}\,\sigma \,, \ee \be \label{gcom}
\left [ a^\alpha_i\,,\,a^\beta_j \right ]= \epsilon^{\alpha\,\beta} A_{ij}\,,
\ee where $i,j=1,\,...,\,N$, $\ \alpha ,\beta=0,1$, $\
\epsilon^{\alpha\,\beta}$ =$- \epsilon^{\beta \,\alpha},$ $\
\epsilon^{0\,1}=1$ and \be \label{A} A_{ij}=\delta_{ij} +\nu\tilde{A}_{ij}
\,,\qquad \tilde{A}_{ij}= \delta_{ij}\sum_{l=1}^N K_{il} - K_{ij}\,. \ee Here
$K_{ij}\in S_N$ with $i,\,j\,=\,1,\,...\,,\,N\,,$ $i \neq j$, are the
elementary permutations $i \leftrightarrow j$ satisfying the relations $$
K_{ij}=K_{ji},\ K_{ij}\,K_{ij}=1,\ K_{ij}K_{jl}= K_{jl}K_{li}= K_{li}K_{ij}
$$ for $i\neq j \neq l\neq i$ and $$ K_{ij}\,K_{kl}=K_{kl}\,K_{ij} $$ if
$i,\,j,\,k,\,l$ are pairwise different. Note that in this paper repeated
Latin indices $i,j,k,\ldots$ do not imply summation.

The defining relations (\ref{begin})-(\ref{A}) are consistent. In particular,
the Jacobi identities \be \label{jac} [a^\alpha_i\,,[a^\beta_j \,,a^\gamma_k
]]+ [a^\beta_j\,[a^\gamma_k\,, a^\alpha_i\,]]
+[a^\gamma_k\,,[a^\alpha_i\,,a^\beta_j ]]=0 \ee are satisfied.

An important property of $SH_N (\nu )$ which allows one to solve the Calogero
model \cite{4} is that this algebra possesses inner $sl_2$ automorphisms with
the generators \be \label {sl2} T^{\alpha\beta}= \frac 1 2 \sum _{i=1}^N
(a^{\alpha}_i a^{\beta}_i + a^{\beta}_i a^{\alpha}_i )\,, \ee \be \label
{csl2} [T^{\alpha\beta}, T^{\gamma\delta}]= \epsilon^{\alpha\gamma}
T^{\beta\delta} +\epsilon^{\alpha\delta} T^{\beta\gamma} +
\epsilon^{\beta\gamma} T^{\alpha\delta} +\epsilon^{\beta\delta}
T^{\alpha\gamma}\,, \ee which act on the generating elements $a^\alpha_i$ as
on $sl_2$ vectors \be \label {sl2vec} \left [ T^{\alpha\beta},\,a^\gamma_i
\right ]= \epsilon^{\alpha\gamma} a^\beta_i +\epsilon^{\beta\gamma}
a^\alpha_i \,. \ee Calogero Hamiltonian is identified with the Cartan element
$T^{01}$ which according to (\ref {sl2vec}) induces $Z$ - gradation of $SH_N
(\nu )$. The latter property allows one \cite{2} to construct wave functions
via the standard Fock procedure with the Fock vacuum $ |0\rangle$ such that
$a_i^0 |0\rangle$=0. Thus, the elements $a_i^\alpha$ serve as generalized
oscillators underlying the Calogero problem. The concrete realization of
these oscillators in terms of Dunkl differential-difference operators
\cite{6} was presented in \cite{1,2}.

These properties characterize the algebra $SH_N (\nu)$ as a natural
generalization of the Heisenberg-Weyl algebra, the associative algebra of
harmonic oscillators. Since the Lie algebra of quantum operators in the
harmonic oscillator problem can be identified with the $W_{1+\infty}$ algebra
\cite{7}, the Lie (super)algebras constructed from $SH_N (\nu )$ with the aid
of supercommutators give rise to a class of the $W_{1+\infty}$ - type
algebras which have been denoted as $W_{N,\infty} (\nu )$ in \cite{8} where
it was shown that all these algebras contain the Virasoro algebra as a
subalgebra. The latter observation indicates that the algebras $SH_N (\nu )$
and $W_{N,\infty} (\nu )$ can be related to conformal models as well as to
other classes of models in the range from quantum Hall effect \cite{9} till
higher-spin gauge theories \cite{10} and KP hierarchy \cite{11} where
$W_{\infty}$ - type algebras prove to be important. An additional argument in
favor of the relationship of \A to the quantum Hall effect is due to the
known fact that the Calogero model can be interpreted as a one-dimensional
reduction of the full anyonic problem \cite{12, 3}.

For lower values of $N$, a nature of $SH_N (\nu )$ is rather well understood.
$SH_1 (\nu )$ is the ordinary Heisenberg-Weyl algebra (since $\nu$ -
dependence is artificial in this case we will use the notation $SH_1$).
Properties of this algebra are very well studied (see {\it e.g.} \cite{13}).
Note that since the center of mass coordinates $1/N \sum_{i=1}^{N}
a_i^\alpha$ decouple from everything else in the defining relations
(\ref{begin})-(\ref{A}), the associative algebra $SH_N (\nu )$ has the
structure $SH_N (\nu )$= $SH_1 \otimes SH^\prime_N (\nu )$ where, by
definition, $SH^\prime_N (\nu )$ is the algebra of elements depending only on
the relative coordinates $a_i^\alpha - a_j^\alpha$.

The properties of $SH^\prime_2 (\nu )$ are well studied too \cite{14}. The
algebra $SH^\prime_2 (\nu )$ is defined by the relations \be \label{N2}
[a^\alpha ,a^\beta ]=\epsilon^{\alpha\beta}(1+2\nu K)\,, \n \ee where $K$ is
the only nontrivial element of $S_2$ while $a^\alpha$ are the relative motion
oscillators. For the particular case of $\nu=0$ one recovers the algebra
$SH_1$ in the sector of the $K$ independent elements.

In \cite{14} it was shown that $SH^\prime_2 (\nu )$ admits a unique
supertrace operation defined by the simple formula \be \label{str2}
str(1)=1\,,\qquad str(K)=-2\nu\,, \qquad str (W)=str (WK)=0 \n \ee for any
polynomial $W\in SH^\prime_2$ of the form \be \label{W} W=\sum_{n=1}^{\infty}
W_{\alpha_1 \ldots \alpha_n } a^{\alpha_1} \ldots a^{\alpha_n} \n \ee with
arbitrary totally symmetric multispinors $W_{\alpha_1 \ldots \alpha_n }$.
For the particular case of $\nu=0$ one recovers the supertrace on $SH_1$.

Furthermore it was shown in \cite{14} by explicit evaluation of the invariant
bilinear form $B(x,\,y )\stackrel {def}{=} str(xy)$ that for $\nu = l+
\frac{1}{2}$ ($l$ is any integer) $SH^\prime_2 (\nu )$ reduces to
finite-dimensional matrix algebras up to some infinite-dimensional ideals
${\cal I}$ which decouple from everything under the supertrace operation
(\ref{str2}), \ie $str(xy)=0$, $\forall x \in {\cal I}$.

In \cite{15} it was then observed that $SH^\prime_2 (\nu )$ is isomorphic to
the factor of the enveloping algebra $U(osp(1;2))$ of $osp(1;2)$ over its
ideal generated by the quadratic Casimir operator $C_2$ by factoring out all
elements of the form $(C_2-c_2) U(osp(1;2))$ where $c_2=\frac 1 {16}
(4\nu^2-1)$ is an arbitrary constant. In its turn this observation clarified
the origin of the ideals of $SH^\prime_2 (\nu )$ at $\nu=l+\half$ as
corresponding to the finite-dimensional representations of $osp(1;2)$.

Although the algebra $SH_N (\nu )$ is getting interesting applications for
any $N$, till now understanding of its algebraic properties for $N>2$ is far
from being satisfactory. In particular there is no interpretation of $SH_N
(\nu )$ in terms of enveloping algebras of finite-dimensional superalgebras
and nothing is known about ideals of $SH_N (\nu )$ which information is very
important in applications.

In this paper we analyze the existence of the invariant supertrace operation
on $SH_N (\nu)$ \ie such a complex valued linear function $str(f)$ on $SH_N
(\nu )$ that \be \label{scom} str\left (\left [f\,,g\right \}\right
)=0\,,\qquad\forall f,g\in SH_N (\nu) \ee with the convention that \be
\label{com} \left [f\,,g\right \}=fg-(-1)^{\pi(f) \pi (g)}gf\,, \ee where the
parity $\pi$ in $SH_N(\nu )$ is defined as follows: \be \label {m0} \pi
(a^\alpha_i)=1\,,\ \ \pi (K_{ij})=0\,. \ee Let us note that an attempt to
define differently graded traces like, $e.g.$, an ordinary trace ($\pi\equiv
0$) unlikely leads to interesting results.

Knowledge of the supertrace operations on \A is useful in various respects.
One of the most important applications of the supertrace is that it gives
rise to $n$-linear invariant forms \be \label{form0} str(a_1 a_2 ...a_n ) \n
\ee that allows one to work with the algebra essentially in the same way as
with the ordinary finite-dimensional matrix algebras and, for example,
construct Lagrangians when working with dynamical theories based on
$SH_N(\nu)$. Another useful property is that since null vectors of any
invariant bilinear form span a both-side ideal of the algebra, this gives a
powerful device for investigating ideals which decouple from everything under
the supertrace operation as it happens in $SH_2 (\nu )$ for half-integer
$\nu$. It is also worth mentioning that having an explicit form of the
trilinear form in one or another basis is practically equivalent to defining
a star-product law in the algebra.

An important motivation for the analysis of the supertraces of $SH_N (\nu )$
is due to its deep relationship with the analysis of the representations of
this algebra, which in its turn gets applications to the analysis of the wave
functions of the Calogero model. For example, given representation of $SH_N
(\nu )$, one can speculate that it induces some supertrace on this algebra as
(appropriately regularized) supertrace of (infinite) representation matrices.
When the corresponding bilinear form degenerates this would imply that the
representation becomes reducible.

As we show, the situation for $SH_N (\nu )$ is very interesting since
starting from $N=3$ it admits more than one independent supertrace in
contrast to the cases of $N=1$ and $N=2$. This fact is in agreement with the
results of \cite{5} where it was shown that there exist many inequivalent
lowest-weight type representations of $SH_N (\nu )$ for higher $N$ (these
representations are classified according to the representations of $S_N$.)
Another important consequence of this phenomenon is that the Lie
superalgebras $W_{N,\infty}(\nu)$ are not simple while appropriate their
simple subalgebras possess non-trivial outer automorphisms.

The paper is organized as follows. In Section \ref{sec2} we analyze
consequences of $S_N$ and $sl_2$ automorphisms of $SH_N(\nu)$. In Section
\ref{sec3} we discuss general properties of the supertraces and consequences
of the existence of several independent supertraces. In Section \ref{sec4} we
study the restrictions on supertraces of the group algebra of $S_N$
considered as a subalgebra of $SH_N (\nu )$, which follow from the defining
relations of $SH_N (\nu )$. These restrictions are called ground level
conditions ({\it GLC}). They play a fundamental role in the problem since as
we show in Section \ref{sec5} every solution of {\it GLC} admits a unique
extension to some supertrace on $SH_N(\nu)$. In Appendix \ref{appa} it is
shown that the number of independent supertraces on $SH_N (\nu )$ equals to
the number of partitions of $N$ into a sum of odd positive integers. Some
technical details of the proof of Section \ref{sec5} are collected in
Appendices \ref{appb} and \ref{appc}.

\section{Finite-Dimensional Groups of Automorphisms}\label{sec2}

The group algebra of $S_N$ is the finite-dimensional subalgebra of $SH_N (\nu
)$. The elements $\sigma \in S_N$ induce inner automorphisms of $SH_N(\nu)$.
It is well known, that any $\sigma \in S_N$ can be expanded into a product of
pairwise commuting cycles \be \label{dec} \sigma=c_1 c_2 c_3\, ...\,c_t\,,
\ee where $c_{\fr w}$, $ {\fr w}=1, \dots, t$, are cyclic permutations acting
on distinct subsets of values of indices $i$. For example, a cycle which acts
on the first $s$ indices as $1 \rightarrow 2 \rightarrow \,...\, \rightarrow
s \rightarrow 1$ has the form \be \label{ex} c=K_{12}K_{23}\,...\,
K_{(s-1)\,s}\,. \ee We use the notation $|c|$ for the length of the cycle
$c$. For the cycle (\ref{ex}), $|c|=s$. We take a convention that the cycles
of unit length are associated with all values of $i$ such that $\sigma
(i)=i$, so that the relation $\sum_{\fr w} |c_{\fr w}| = N$ is true.

Given permutation $\sigma \in S_N $, we introduce a new set of basis elements
${\fr B}_\sigma$=$\{b^I\}$ instead of $\{a^\alpha_i\}$ in the following way.
For every cycle $c_{\fr w}$ in the decomposition (\ref{dec}) ($ {\fr
w}=1,\,...\,,\,t $), let us fix some index $l_{\fr w}$, which belongs to the
subset associated with the cycle $c_{\fr w}$. The basis elements
$b^\alpha_{{\fr w}j}$, $j=1,\,...\,,|c_{\fr w}|$, which realize 1-dimensional
representations of the commutative cyclic group generated by $c_{\fr w}$,
have the form \be \label {a} b^\alpha_{{\fr w}j} =\frac 1 {\sqrt{|c_{\fr
w}|}} \sum _{k=1}^{|c_{\fr w} |} (\lambda_{\fr w})^{jk} a^{\alpha}_{l({\fr
w},k)}\,, \ee where $l({\fr w},k)=c_{\fr w}^{-k} (l_{\fr w})$ and
\bee\label{lambda} \lambda_{\fr w}=exp(2\pi i /|c_{\fr w}|). \eee

{}From the definition (\ref{a}) it follows that \be\label{eig} c_{\fr w}
b^\alpha_{{\fr w}j} = (\lambda_{\fr w})^j b^\alpha_{{\fr w}j} c_{\fr w}\,, \n
\ee \be\label{eignext} c_{\fr w} b^\alpha_{{\fr n}j} = b^\alpha_{{\fr n}j}
 c_{\fr w}\mbox{ , for } {\fr n}\neq {\fr w} \n \ee and therefore
\be\label{eigs} \sigma b^\alpha_{{\fr w}j} = (\lambda_{\fr w}
 )^jb^\alpha_{{\fr w}j} \sigma\,. \ee

In what follows, instead of writing $b^\alpha_{{\fr w}j}$ we use the notation
$b^I$ with the label $I$ accounting for the full information about the index
$\alpha$, the index ${\fr w}$ enumerating cycles in (\ref{dec}), and the
index $j$ which enumerates various elements $b^\alpha_{{\fr w}j}$ related to
the cycle $c_{\fr w}$, \ie $I$ ($I = 1,\,...,\,2N$) enumerates all possible
triples $\{\alpha,{\fr w},j\}$. We denote the index $\alpha$, the cycle and
the eigenvalue in (\ref{eig}) corresponding to some fixed index $I$ as
$\alpha (I)$, $ c(I),\ $ and $\lambda_I=(\lambda_{\fr w})^j$, respectively.
The notation $\sigma(I)=\sigma_0$ implies that $b^I \in {\fr B}_{\sigma_0}$.
${\fr B}_{\bf 1}$ is the original basis of the generating elements
$a_i^\alpha$ (here ${\bf 1}$ is the unit permutation).

Let ${\fr M}(\sigma)$ be the matrix which maps ${\fr B}_{\bf
1}\longrightarrow {\fr B}_\sigma$ in accordance with (\ref{a}),
\bee\label{frm} b^I=\sum_{i,\alpha} {\fr M}_{i\alpha}^I(\sigma)\,
a^\alpha_i\,. \eee Obviously this mapping is invertible. Using the matrix
notations one can rewrite (\ref{eigs}) as \bee\label{eigmat} \sigma b^I
\sigma^{-1}=\sum_{J=1}^{2N} \Lambda^I_J(\sigma)\, b^J\,,\ \ \forall b^I \in
 {\fr B}_\sigma\,, \eee where $\Lambda_I^J(\sigma)=\delta_I^J \lambda_I$.

Every polynomial in $SH_N(\nu)$ can be expanded into a sum of monomials of
the form \be \label{[} b^{I_1} b^{I_2}...\,b^{I_s}\sigma\,, \ee where all
$\sigma (I_k ) =\sigma$. Every monomial of this form realizes some
one-dimensional representation of the Abelian group generated by all cycles
$c_{\fr w}$ in the decomposition (\ref{dec}).

The commutation relations for the generating elements $b^I$ follow from
(\ref{gcom}) and (\ref{A}) \be\label{f} \left [ b^I,\,b^J\right]= F^{IJ}=
{\cal C}^{IJ} + \nu f^{IJ}\,, \n \ee where \be\label{calC} {\cal
C}^{IJ}=\epsilon^{\alpha (I) \alpha (J)} \delta_{c(I) c(J)} \delta_{\lambda_I
\lambda_J^{-1}} \ee and \be \label{struc} f^{IJ}=\sum_{i,j,\alpha ,\beta}{\fr
M}^I_{i\alpha} (\sigma ) {\fr M}^J_{j\beta} (\sigma )
\epsilon^{\alpha\beta}\tilde{A}_{ij}. \ee

The indices $I,J$ are raised and lowered with the aid of the symplectic form
$ {\cal C}^{IJ}$ \be\label{rise} \mu^I=\sum_J{\cal C}^{IJ}\mu_J\,,\qquad
\mu_I=\sum_J\mu^J {\cal C}_{JI}\,; \qquad \sum_M{\cal C}_{IM}{\cal
C}^{MJ}=-\delta_I^J\,. \ee Note that the elements $b^I$ are normalized in
(\ref{a}) in such a way that the $\nu$-independent part in (\ref{f}) has the
form (\ref{calC}).

Another important finite-dimensional algebra of inner automorphisms of $SH_N
(\nu )$ is the $sl_2$ algebra which acts on the indices $\alpha$. It is
spanned by the $S_N$-invariant second-order polynomials (\ref {sl2}).
Evidently, $SH_N(\nu)$ decomposes into the infinite direct sum of only
finite-dimensional irreducible representations of this $sl_2$ spanned by
various homogeneous polynomials (\ref{[}).

{}From the defining relations (\ref{begin})-(\ref{A}) it follows that $SH_N
(\nu )$ is $Z_2$ - graded with respect to the automorphism \bee\label{auto}
f(a_j^\alpha)=-a_j^\alpha\,, \qquad f(K_{ij} )=K_{ij} \eee which gives rise
to the parity $\pi$ (\ref{m0}). In applications to higher-spin models, this
automorphism distinguishes between bosons and fermions.

The algebra $SH_N(\nu )$ admits the antiautomorphism $\rho$, \be \label{ant}
\rho (a^\alpha_k )= ia^\alpha_k\,,\qquad \rho(K_{ij}) = K_{ij}\,, \n \ee
which leaves invariant the basic relations (\ref{begin})-(\ref{A}) provided
that an order of operators is reversed according to the defining property of
antiautomorphisms: $\rho (AB)$= $\rho (B)\rho (A)$. {}From (\ref{dec}),
(\ref{ex}) and (\ref{eigs}) it follows that \be \label{Ant} \rho (\sigma )=
\sigma^{-1}\,,\qquad \rho (b^I)= ib^J\,, \n \ee where $J$ is related to $I$
in such a way that $\alpha(J)=\alpha(I)$, $\sigma (J) = (\sigma (I))^{-1}$,
$c(J)=(c(I))^{-1}$ and $\lambda_J = \lambda^{-1}_I$. Note that in higher-spin
theories the counterpart of $\rho$ distinguishes between odd and even spins
\cite{16}.

\section{General Properties of Supertrace}\label{sec3}

In this section we summarize some general properties to be respected by any
supertrace in $SH_N (\nu )$.

Let $A$ be an arbitrary associative $Z_2$ graded algebra with the parity
function $\pi(x) =0$ or $1$. Suppose that $A$ admits some supertrace
operations $str_p$ where the label $p$ enumerates different nontrivial
supertraces. We call a supertrace $str $ even (odd) if $str(x)=0 \ \forall x
\in A$ such that $\pi(x)=1(0)$. Let $T_A$ be a linear space of supertraces on
$A$. We say that $dim\,T_A$ is the number of supertraces on $A$.

Given parity-preserving (anti)automorphism $\tau$ and supertrace operation
$str$ on $A$, $str (\tau (x))$ is some supertrace as well. For inner
automorphisms $\tau$ ($\tau (x)= pxp^{-1},\ \pi (p)=0$) it follows from the
defining property of the supertrace that $str (\tau (x))$=$str (x)$. Thus,
$T_A$ forms a representation of the factor-group of the parity preserving
 automorphisms and antiautomorphisms of $A$ over the normal subgroup of the
inner automorphisms of $A$. Applying this fact to the original parity
automorphism $(-1)^\pi$ one concludes that $T_A$ can always be decomposed
into a direct sum of subspaces of even and odd supertraces, $T_A =T_A^0
\oplus T_A^1$ and that $T_A^1 =0$ if the parity automorphism is inner.

In the sequel we only consider the case where $dim\,T_A \ <\infty$ and there
are no nontrivial odd supertraces. Let $A=A_1 \otimes A_2$ with the
associative algebras $A_1$ and $A_2$ endowed with some even supertrace
operations $t_1$ and $t_2$, respectively. The supertrace on $A$ can be
defined by setting $str(a_1 \otimes a_2 )$= $t_1(a_1)$ $t_2(a_2 )$, $\forall
a_1 \in A_1$, $\forall a_2 \in A_2$. As a result, one concludes that $\ T_A$=
$T_{A_1}\otimes T_{A_2}$. In the case of $SH_N (\nu )$ one thus can always
separate out a contribution of the center of mass coordinates as an overall
factor ($SH_1$ admits the unique supertrace).

If $A$ is finite-dimensional then the existence of two different supertraces
indicates that $A$ admits non-trivial both-side ideals. Actually,
con\-si\-der the bilinear form $B(f,g)$ = $\alpha_1 str_1 (fg) + \alpha_2
str_2 (fg)$ with arbitrary parameters $\alpha_1,\ \alpha_2\in {\open C} $ and
elements $f,\ g \in A$. The determinant of this bilinear form is some
polynomial of $\alpha_{1}$ and $\alpha_2$. Therefore it vanishes for certain
ratios $\alpha_{1}/\alpha_{2}$ or $\alpha_{2}/\alpha_{1}$ according to the
central theorem of algebra. Thus, for these values of the parameters the
bilinear form $B$ degenerates and admits non-trivial null vectors $x$,
$B(x,g)$=0, $\forall g\in A$. It is easy to see that the linear space ${\cal
I}$ of all null vectors $x$ is some both-side ideal of $A$. For
infinite-dimensional algebras the existence of several supertraces does not
necessarily imply the existence of ideals. As mentioned in introduction the
existence of several supertrace operations may be related to the existence of
inequivalent representations. Also it is worth mentioning that for the case
of infinite-dimensional algebras and representations under investigation it
can be difficult to use the standard (\ie matrixwise) definition of the
supertrace. In this situation the formal definition of the supertraces on the
algebra we implement in this paper is the only rigorous one.

Let $l_A$ be the Lie superalgebra which is isomorphic to $A$ as a linear
space and is endowed with the product law (\ref{com}). It contains the
subalgebra $sl_A \in l_A $ spanned by elements $g$ such that $str_p (g)=0$
for all $p$. Evidently $sl_A$ forms the ideal of $l_A$. The factor algebra
$t_A$=$l_A$/$sl_A$ is a commutative Lie algebra isomorphic to $T_A^*$ as a
linear space. Elements of $t_A$ different from the unit element of $A$ (which
exist if $dim\,T_A >1$) can induce outer automorphisms of $sl_A$. Let us note
that it is this $sl_A$ Lie superalgebra which usually has physical
applications. For the case of $SH_N (\nu )$ under consideration the algebra
$l_{SH_N(\nu)}$ is identified with the algebra $W_{N,\infty }(\nu )$
introduced in \cite{8}. We therefore conclude that these algebras are not
simple for $N>2$ because it is shown below that $SH_N (\nu )$ admits several
supertraces for $N>2$. Instead one can consider the algebras $sW_{N,\infty}
(\nu )$.

Let $l_A$ contain some subalgebra ${\cal L} $ such that $A$ decomposes into a
direct sum of irreducible representations of ${\cal L}$ with respect to the
adjoint action of ${\cal L}$ on $A$ via supercommutators. Then, only trivial
representations of ${\cal L}$ can contribute to any supertrace on $A$.
Actually, consider some non-trivial irreducible representation $R$ of ${\cal
L}$. Any $r\in R$ can be represented as \be \label{ir} r=\sum_j [l_j ,r_j
\}\,,\qquad l_j \in {\cal L},\quad r_j \in R\, \ee since elements of the form
(\ref{ir}) span the invariant subspace in $R$. {}From (\ref{scom}) it follows
then that $str(r)=0\,,\forall r\in R$.

{}From the definition of the supertrace it follows that \be str(a_1 a_2
)+str(a_2 a_1 )=0 \n \ee for arbitrary odd elements $a_1$ and $a_2 $ of $A$.
A simple consequence of this relation is that \be \label{odd} str(a_1
a_2\ldots a_n +a_2 \ldots a_n a_1 +\ldots + a_n a_1 \ldots a_{n-1} ) =0 \n
\ee is true for an arbitrary even $n$ if all $a_i$ are some odd elements of
$A$. Since we assume that the supertrace is even (\ref{odd}) is true for any
$n$. This simple property turns out to be practically useful because, when
odd generating elements are subject to some commutation relations with the
right hand sides expressed via even generating elements like in (\ref{gcom}),
it often allows one to reduce evaluation of the supertrace of a degree-$n$
polynomial of $a_i$ to supertraces of lower degree polynomials.

Another useful property is that in order to show that the characteristic
property of the supertrace (\ref{scom}) is true for any $x,g\,\in A$, it
suffices to show this for a particular case where $x$ is arbitrary while $g$
is an arbitrary generating element of some fixed system of generating
elements. Then (\ref{scom}) for general $x$ and $g$ will follow from the
properties that $A$ is associative and $str$ is linear. For the particular
case of $SH_N (\nu )$ this means that it is enough to set either
$g=a_i^\alpha$ or $g=K_{ij}$.

Let us now turn to some specific properties of $SH_N (\nu)$ as a particular
realization of $A$.

By identifying ${\cal L}$ with $sl_2$ (\ref{sl2}) and taking into account
that $SH_N(\nu )$ decomposes into a direct sum of irreducible
finite-dimensional representations of $sl_2$, one arrives at the following

\noindent {\it {\bf Lemma 1}: $str(x)$ can be different from zero only when
$x$ is $sl_2$-singlet, \ie $[T^{\alpha\beta}, x]=0$.}

\noindent {\it Corollary}: Any supertrace on $SH_N (\nu )$ is even.

Analogously one deduces consequences of the $S_N$ symmetry. In particular,
one proves

\noindent {\it {\bf Lemma 2}: Given $c\in S_N$ such that $cF=\mu Fc$ for some
element $F$ and any constant $\mu\neq 1$, $str(F)=0\,$. Given monomial
$F=b^{I_1}b^{I_2}\,...\,b^{I_s} \sigma$ with $b^{I_k}\in {\fr B}_\sigma $ and
a cycle $c_0$ in the decomposition (\ref{dec}) of $\sigma$ one concludes that
$str(F)=0\,$ if $\,\prod_{k:\,c(I_k)=c_0} \lambda_{I_k} \neq 1 $ where
$\lambda_{I_k}$ are the eigenvalues (\ref{eigs}) of $b^{I_k}$.}

\section{Ground Level Conditions}\label{sec4} Let us analyze restrictions on
a form of $str(a)$, $a\in S_N$, which follow from the defining relations of
$SH_N (\nu )$.

Firstly, we describe supertraces on the group algebra of $S_N$. Let some
permutation $\sigma$ decomposes into $n_1$ cycles of length $1$, $n_2$ cycles
of length $2$, ... and $n_N$ cycles of length $N$. The non-negative integers
$n_k$ satisfy the relation \be\label{m} \sum_{k=1}^N kn_k =N \ee and fix
$\sigma$ up to some conjugation $\sigma \rightarrow \tau \sigma \tau^{-1}$,
$\tau \in S_N$. Thus \be\label{varphi} str(\sigma)=\varphi
(n_1,n_2,\,...\,,n_N)\,, \ee where $\varphi (n_1,n_2,\,...\,,n_N) $ is an
arbitrary function. Obviously the linear space of invariant functions on
$S_N$ (\ie such that $f(\tau\sigma\tau^{-1})=f(\sigma)$) coincides with the
linear space of supertraces on the group algebra of $S_N$. Therefore, the
dimension of the linear space of supertraces is equal to the number $p(N)$ of
independent solutions of (\ref{m}), the number of conjugacy classes of $S_N$.
One can introduce the generating function for $p(N)$ as
$P(q)=\sum_{n=0}^\infty p(n) q^n$=$\prod_{k=1}^\infty \frac 1 {(1-q^k)}$.
The properties of this generating function and of the quantities $p(N)$ are
discussed in details \eg in \cite{17}.

According to the general argument of the previous section, the existence of
several independent traces implies that the group algebra of $S_N$ must have
some ideals. Indeed it can be shown to decompose into a direct sum of matrix
algebras $Mat_n$.

Since the group algebra of $S_N$ is embedded into $SH_N (\nu)\,$ some
additional restrictions on the functions $\varphi (n_1,n_2,\,...,\,n_N) $
follow from (\ref{scom}) and the defining relations (\ref{gcom})-(\ref{A}) of
$SH_N (\nu )$. Actually, consider some elements $b^I$ such that
$\lambda_I=-1$. Then, one finds from (\ref{scom}) and (\ref{eigs}) that $str
\left ( b^I b^J \sigma \right )$= $ - str \left ( b^J \sigma b^I\right )$= $
str \left ( b^J b^I \sigma \right )$ and therefore \be\label{ccc} str \left (
[ b^I, b^{J}] \sigma \right ) =0\,. \ee Since these conditions restrict
supertraces of degree-0 polynomials of $a^\alpha_i$ we call them ground level
conditions ({\it GLC}). Thus for every permutation $\sigma$ and any even
integer $2k$ such that there exists some cycle $c$ of length $|c|=2k$ in the
decomposition (\ref{dec}) we have {\it GLC} (\ref{ccc}) with $b^I$ such that
$c(I)=c$. Note however that if $\lambda_J\neq -1$ or $c(J) \neq c(I)$ then
the relation $str ([ b^I, b^J] \sigma ) =0$ is trivially satisfied as a
consequence of {\it Lemma 2}.

It is convenient to rewrite {\it GLC} in the following form \bee \label{eqss}
str(c_0\sigma_0)= -str\left(\ig ([b^0_{0k},b^1_{0k}]-1\ig )c_0\sigma_0 \right
), \eee where $c_0$ is any cycle of even length $2k$ in the decomposition of
the permutation $\sigma=c_0\sigma_0$ and $b_{0k}^\alpha$ is the corresponding
variable (\ref{a}) with $(\lambda_0)^k=-1$, \ie $c_0 b_{0k}^\alpha=
-b_{0k}^\alpha c_0\,,\ $ $\sigma_0 b_{0k}^\alpha= b_{0k}^\alpha\sigma_0$ and
$\lambda_0=exp(2\pi i/|c_0|)$.

To work out the explicit form of the restrictions on the functions $\varphi
(n_1,n_2,\,...\,n_N) $ which follow from {\it GLC} one has to use the
following simple facts from the theory of the symmetric group:

\noindent {\it {\bf Lemma 3}: Let $c_1$ and $c_2$ be two distinct cycles in
the decomposition (\ref{dec}). Let indices $i_1$ and $i_2$ belong to the
subsets of indices associated with the cycles $c_1$ and $c_2$, respectively.
Then the permutation $c= c_1 c_2 K_{i_1\, i_2}$ is a cycle of length $|c|=
|c_1| + |c_2| $.}

\noindent {\it {\bf Lemma 4}: Given cyclic permutation $c \in S_N$, let
$i\neq j$ be two indices such that $c^k (i) = j$, where $k$ is some positive
integer, $k<|c|$. Then $c K_{ij} = c_1 c_2 $ where $c_{1,2}$ are some
non-coinciding mutually commuting cycles such that $|c_1|=k$ and $|c_2|=
|c|-k$.}

Using the definition (\ref{a}), the commutation relations
(\ref{begin})-(\ref{A}) and {\it Lemmas 3} and {\it 4} one reduces {\it GLC}
 to the following system of equations: \bee\label {mm} &{}& n_{2k}\varphi
(n_1,\,...\,,n_{2k},\,...\,,n_N) \nn &{}&= -\nu n_{2k} \bigg ( 2 \sum_{s\neq
k,\,s=1}^{2k-1} O_s \varphi (n_1,\,...\,,n_s+1,\,...\,,
n_{2k-s}+1,\,...\,,n_{2k}-1,\,...\,,n_N) \nn &{}& + 2 O_k \varphi
(n_1,\,...\,,n_k+2,\,...\,,n_{2k}-1,\,...\,,n_N) \nn &{}&+ \sum_{s\neq
2k;\,s=1}^N sn_s \varphi (n_1,\,...\,,n_s-1,\,...\,,n_{2k}-1,\,...
\,,n_{2k+s}+1,\,...\,,n_N) \nn &{}&+ 2k(n_{2k}-1)\varphi
(n_1,\,...\,,n_{2k}-2,\,...\,,n_{4k}+1,\,...\,,n_N) \bigg ) \eee where
$O_k=0$ for $k$ even and $O_k=1$ for $k$ odd.

Let us note that by virtue of the substitution \be \label{scal} \varphi
(n_1,\,\ldots \,,n_N)= \nu^{E(\sigma)} {\tilde \varphi} (n_1,\,\ldots
\,,n_N)\,, \ee where $E(\sigma)$ is the number of cycles of even length in
the decomposition of $\sigma$ (\ref{dec})\,, \ie \be E(\sigma)=n_2+n_4+\,...
\ee one can get rid of the explicit dependence of $\nu$ from {\it GLC}
(\ref{mm}). As a result, there are two distinguishing cases, $\nu=0$ and $\nu
\neq 0$.

For lower $N$ the conditions (\ref {mm}) take the form \be
\varphi(0,1)+2\nu\varphi(2,0)=0 \n \ee for $N=2$ ({\it cf.} (\ref{str2})),
\be \varphi(1,1,0)+2\nu\varphi(3,0,0)+\nu\varphi(0,0,1)=0 \n \ee for $N=3$
and \begin{eqnarray}
\varphi(2,1,0,0)+2\nu\varphi(4,0,0,0)+2\nu\varphi(1,0,1,0)&=&0 \nn
\varphi(0,2,0,0)+2\nu\varphi(2,1,0,0)+2\nu\varphi(0,0,0,1)&=&0 \nn
\varphi(0,0,0,1)+4\nu\varphi(1,0,1,0)&=&0 \nonumber \end{eqnarray} for $N=4$.
As a result one finds 1-parametric families of solutions for $N=1$ and $N=2$
and 2-parametric families of solutions for $N=3$ and $N=4$.

Let $G_N$ be the number of independent solutions of (\ref{mm}). As we show in
the next section $G_N$=$dimT_{SH_N (\nu )}$ for all $\nu$. In other words all
other conditions on the supertrace do not impose any restrictions on the
functions $\varphi (n_1, \ldots ,n_N )$ but merely express supertraces of
higher order polynomials of $a_i^\alpha$ in terms of $\varphi (n_1, \ldots
,n_N )$.

In the Appendix \ref{appa} we prove the following

\noindent {\it {\bf Theorem 1}: $G_N=q(N)$ where $q(N)$ is a number of
partitions of $N$ into a sum of odd positive integers, \ie the number of the
solutions of the equation $\sum_{k=0}^\infty (2k+1) n_{k}=N$ for non-negative
integers $n_i$.}

One can guess this result from the particular case of $\nu=0$ where {\it GLC}
tell us that $\varphi(n_1,\,...\,,\,n_N)$ can be nonvanishing (and arbitrary)
only when all $n_{2k}=0$. Interestingly enough, $G_N$ remains the same for
$\nu\neq 0$.

\section{Supertrace for General Elements}\label{sec5}

In this section we prove

\noindent {\it {\bf Theorem 2}: $dimT_{SH_N (\nu )}$=$G_N$ where $G_N$ is the
number of independent solutions of the ground level conditions (\ref{mm}).}

The proof of the {\it Theorem 2} will be given in a constructive way by
virtue of the following double induction procedure:

\noindent {\bf (i)}. Assuming that {\it GLC} are true and $str\{b^I , P_p (a)
\sigma \}=0$ $\forall P_p (a),$ $\sigma$ and $I$ provided that $b^I \in {\fr
B}_\sigma$ and $$ \begin{array}{l} \mbox{$\lambda (I) \neq -1$; $p\leq k\,$
or}\\ \mbox{$\lambda (I) =-1$, $E(\sigma )\leq l$, $p\leq k\,$ or}\\
\mbox{$\lambda (I) =-1$; $p\leq k-2$}\,, \end{array} $$ where $P_p (a)$ is an
arbitrary degree $p$ polynomial of $a_i^{\alpha} $ ($p$ is odd) and $E(\sigma
)$ is the number of cycles of even length in the decomposition (\ref{dec}) of
$\sigma$, one proves that there exists such a unique extension of the
supertrace that the same is true for $l\rightarrow l+1$.

\noindent {\bf (ii)}. Assuming that $str\{b^I , P_p (a) \sigma \}=0$ $\forall
P_p (a)$, $\sigma$ and $b^I$ such that $\sigma (I)=\sigma$, $p\leq k$ one
proves that there exists such a unique extension of the supertrace that the
assumption {\bf (i)} is true for $k\rightarrow k+2$ and $l=0$.

As a result this inductive procedure extends uniquely any solution of {\it
GLC} to some supertrace on the whole $SH_N (\nu )$. (Let us remind ourselves
that the supertrace of any odd element of $SH_N (\nu )$ is trivially zero by
$sl_2$ invariance).

The inductive proof of the {\it Theorem 2} is based on the $S_N$ covariance
of the whole setting and the following important

\noindent {\it {\bf Lemma 5}: Given permutation $\sigma$ which has $E(\sigma
)$ cycles of even length in the decomposition (\ref{dec}), the quantity
$f^{IJ}\sigma $ for $\sigma (I)=\sigma (J)=\sigma $ and $\lambda_I=\lambda_J
=-1 $ can be uniquely expanded as $ f^{IJ}\sigma =\sum_q {\alpha}_q \sigma_q
$ where ${\alpha}_q$ are some coefficients and $E(\sigma_q )=E(\sigma )-1$
$\forall q$.}

{\it Lemma 5} is a simple consequence of the particular form of the structure
coefficients $f^{IJ}$ (\ref{struc}) and {\it Lemmas 3} and {\it 4}. The proof
is straightforward. Let us stress that it is {\it Lemma 5} which accounts for
the specific properties of the algebra $SH_N (\nu )$ in the analysis of this
section.

In practice it is convenient to work with the exponential generating
functions \be \label{gf} \Psi_\sigma (\mu )= str\left ( e^S \sigma \right
)\,,\qquad S= \sum_{L=1}^{2N} (\mu_{L } b^{L} )\,, \n \ee where $\sigma$ is
some fixed element of $S_N$, $b^L \in {\fr B}_\sigma $ and $\mu_{L } \in
{\open C}$ are independent parameters. By differentiating over $\mu_{L }$ one
can obtain an arbitrary polynomial of $b^L$ in front of $\sigma$. The
exponential form of the generating functions implies that these polynomials
 are Weyl ordered. In these terms the induction on a degree of polynomials is
equivalent to the induction on a degree of homogeneity in $\mu$ of the power
series expansions of $\Psi_\sigma (\mu )$.

As a consequence of the general properties discussed in the preceding
sections the generating function $\Psi_\sigma (\mu )$ must be invariant under
the $S_N$ similarity transformations \bee\label{S_N}
\Psi_{\tau\sigma\tau^{-1}}(\mu)=\Psi_\sigma (\tilde{\mu})\,, \eee where the
$S_N$ transformed parameters are of the form \bee \label{base}
\tilde{\mu}_I=\sum_J \left({\fr M}(\tau\sigma\tau^{-1}) {\fr
M}^{-1}(\tau)\Lambda^{-1}(\tau){\fr M}(\tau) {\fr M}^{-1}(\sigma)\right)_I^J
{\mu}_J \eee and matrices ${\fr M}(\sigma)$ and $\Lambda(\sigma)$ are defined
in (\ref{frm}) and (\ref{eigmat}).

In accordance with the general argument of Section \ref{sec3} the necessary
and sufficient conditions for the existence of even supertrace are the
$S_N$-covariance conditions (\ref{S_N}) and the condition that
\be\label{start} str\left \{b^L , (exp S ) \sigma \right \}=0\qquad \mbox{for
any $\sigma$ and $L$} \,. \ee To transform (\ref{start}) to an appropriate
form, let us use the following two general relations which are true for
arbitrary operators $X$ and $Y$ and the parameter $\mu \in {\open C}$:
\be\label{r1} Xexp(Y+\mu X)=\f{\p}{\p\mu}exp (Y+\mu X )+ \int \,t_2 \,exp(t_1
(Y+\mu X))[X,Y] exp(t_2 (Y+\mu X))D^1t , \ee \be\label{r2} exp(Y+\mu
X)X=\f{\p}{\p\mu}exp (Y+\mu X )- \int \,t_1 \,exp(t_1 (Y+\mu X))[X,Y] exp(t_2
(Y+\mu X))D^1t \ee with the convention that \be\label{t} D^{n-1}t=\delta (t_1
 +\ldots +t_n -1)\theta (t_1 )\ldots \theta (t_n ) dt_1 \ldots dt_n \,. \ee

The relations (\ref{r1}) and (\ref{r2}) can be derived with the aid of the
partial integration (\eg over $t_1$) and the following formula \be\label{d}
\f{\p}{\p\mu}exp (Y+\mu X ) = \int \, exp(t_1 (Y+\mu X)) X exp(t_2 (Y+\mu
X))D^1 t\,, \ee which can be proven by expanding in power series. The
well-known formula \be \label{r3} [X,exp(Y)]= \int \, exp(t_1 Y)[X,Y] exp(t_2
Y)D^1 t \ee is a consequence of (\ref{r1}) and (\ref{r2}).

With the aid of (\ref{r1}), (\ref{r2}) and (\ref{eigs}) one rewrites
(\ref{start}) as \be\label{nm1} (1+\lambda_L )\f{\p}{\p\mu_L }\Psi_\sigma
(\mu )= \int \,(\lambda_L t_1 -t_2 ) str\ig ( exp (t_1 S)[b^L ,S]\,exp (t_2
S)\sigma\ig )\, D^1 t\,. \ee This condition should be true for any $\sigma$
and $L$ and plays the central role in the analysis of this section.

There are two essentially distinguishing cases, $\lambda_L \neq -1$ and
$\lambda_L =-1$. In the latter case, the equation (\ref{nm1}) takes the form
\be\label{m1} 0=\int \, str\ig ( exp (t_1 S)[b^L ,S]\,exp (t_2 S) \sigma \ig
)D^1 t\,,\qquad \lambda_L =-1\,. \ee

In Appendix \ref{appb} we show by induction that the equations (\ref{nm1})
and (\ref{m1}) are consistent in the following sense \bee\label{c1}
(1+\lambda_K )\f{\p}{\p\mu_K }\int\,(\lambda_L t_1 -t_2 ) str\ig ( exp (t_1
S)[b^L ,S]\,exp (t_2 S) \sigma \ig )D^1 t -(L \leftrightarrow K )=0, & & \\
\lambda_L \neq -1, \ \lambda_K \neq -1 & & \nonumber \eee and \be \label{c2}
(1+\lambda_K )\f{\p}{\p\mu_K }\int \, str\ig ( exp (t_1 S)[b^L ,S]\,exp (t_2
S)\sigma\ig )D^1 t\,=0 ,\qquad \lambda_L=-1. \ee Note that this part of the
proof is quite general and does not depend on a concrete form of the
commutation relations of $a_i^{\alpha}$ in (\ref{gcom}).

By expanding the exponential $e^S$ in (\ref{gf}) into power series in $\mu
_K$ (equivalently $b^K$) one concludes that the equation (\ref{nm1}) uniquely
reconstructs the supertrace of monomials containing $b^K$ with $\lambda_K\neq
-1$ (from now on called regular polynomials) via supertraces of some lower
order polynomials. The consistency conditions (\ref{c1}) and (\ref{c2}) then
guarantee that (\ref{nm1}) does not impose any additional conditions on the
supertraces of lower degree polynomials and allow one to represent the
generating function in the form \bee\label{ex1} \Psi_\sigma &=&
\Phi_\sigma(\mu)\\ &+& \sum_{L:\,\lambda_L \neq -1} \int_0^1 \frac {\mu_L
d\tau} {1+\lambda_L}\int D^1 t\,(\lambda_L t_1 -t_2 ) str\ig ( e^{t_1 (\tau
S^{\prime\prime}+S^\prime)}[b^L ,(\tau S^{\prime\prime}+S^\prime)] \,e^{ t_2
(\tau S^{\prime\prime}+S^\prime)}\sigma\ig )\, , \nonumber \eee where we have
introduced the generating functions $\Phi_\sigma$ for the supertrace of
special polynomials, \ie the polynomials depending only on $b^L$ with
$\lambda_L=-1$, \be\label{gff} \Phi_\sigma (\mu )\stackrel {def}{=} str\left
( e^{ S^\prime} \sigma \right ) = \Psi_\sigma (\mu)\ig |_ {(\mu_I=0\ \forall
I:\ \lambda_I \neq -1)} \n \ee and \be \label{spr} S^\prime = \sum_{L:\,b^L
\in {\fr B}_\sigma,\,\lambda_L=-1} (\mu_{L } b^L); \qquad
 S^{\prime\prime}=S-S^\prime\,. \ee The relation (\ref{ex1}) successively
expresses the supertrace of higher order regular polynomials via the
supertraces of lower order polynomials.

One can see that the arguments above prove effectively the inductive
hypotheses {\bf (i)} and {\bf (ii)} for the particular case where either the
polynomials $P_p (a)$ are regular and/or $\lambda_I \neq -1$. Note that for
this case the induction on the number of cycles of even length {\bf (i)} is
trivial: one simply proves that a power of polynomial can be increased by
two.

Let us now turn to the less trivial case of the special polynomials:
\be\label{startprime} str\left \{b^I , (exp S^\prime ) \sigma \right
\}=0\,,\qquad \lambda_I =-1. \ee Consider the part of $str\left \{b^I , (exp
S^\prime ) \sigma \right \}$ which is of order $k$ in $\mu$ and suppose that
$E(\sigma)=l+1$. According to (\ref{m1}) the conditions (\ref{startprime})
give \be \label{m1prime} 0= \int \, str\left( exp (t_1 S^\prime)[b^I
,S^\prime]\, exp (t_2 S^\prime) \sigma \right) D^1 t\,. \n \ee

Substituting $[b^I ,S^\prime]=\mu^I + \nu \sum_M f^{IM}\mu_M$, where the
quantities $f^{IJ}$ and $\mu^I$ are defined in (\ref{f})-(\ref{rise}), one
can rewrite the equation (\ref{m1prime}) in the form \be\label{formprime}
\mu^I \Phi_\sigma(\mu ) = -\nu \int str\bigg ( exp (t_1 S^\prime)\sum_M
f^{IM}\mu_M\, exp (t_2 S^\prime) \sigma \bigg) D^1 t\,. \ee

Now we use the inductive hypothesis {\bf (i)}. The right hand side of
 (\ref{formprime}) is a supertrace of at most a degree $k-1$ polynomial of
$a^{\alpha}_i$ in the sector of degree $k$ polynomials in $\mu$. Therefore
one can use the inductive hypothesis {\bf (i)} to obtain $$ \int str\ig (
exp(t_1 S^\prime)\sum_M f^{IM}\mu_M\, exp(t_2 S^\prime) \sigma \ig )D^1t =
\int \, str\ig ( exp(t_2 S^\prime)exp(t_1 S^\prime) \sum_M f^{IM}\mu_M \sigma
\ig )D^1 t, $$ where we made use of the simple fact that $str(S^\prime F
\sigma)$ $=$ $ -str(F \sigma S^\prime) $= $str(F S^\prime \sigma)$ due to the
definition of $S^\prime$.

As a result, the inductive hypothesis allows one to transform
(\ref{startprime}) to the following form \be \label{p9} X^I \equiv \mu^I
\Phi_\sigma(\mu ) +\nu str\bigg( exp (S^\prime ) \sum_Mf^{IM}\mu_M\sigma
\bigg)=0 \,. \ee

By differentiating this equation with respect to $\mu_J$ one obtains after
symmetrization \be \label{p10} \f{\p}{\p\mu_J} \left( \mu^I \Phi_\sigma (\mu
)\right) +(I\leftrightarrow J )=-\nu \int str\ig (e^{t_1 S^\prime } b^Je^{t_2
S^\prime } \sum_Mf^{IM}\mu_M \sigma \ig )D^1 t +(I\leftrightarrow J ). \ee

An important point is that the system of equations (\ref{p10}) is equivalent
to the original equations (\ref{p9}) except for the ground level part
$\Phi_\sigma (0)$. This can be easily seen from the simple fact that the
general solution of the system of equations $\f{\p}{\p\mu_J} X^I(\mu) +
\f{\p}{\p\mu_I} X^{J}(\mu) =0 $ for entire functions $X^I(\mu)$ is of the
form $X^I(\mu)=X^I(0)+\sum_{J}c^{IJ}\mu_J$ where $X^I(0)$ and
$c^{JI}$=$-c^{IJ}$ are some constants. The part of (\ref{p9}) linear in $\mu$
is however equivalent to the ground level conditions analyzed in the previous
section. Thus (\ref{p10}) contains all information additional to (\ref{mm}).
For this reason we will from now on analyze the equation (\ref{p10}).

Using again the inductive hypothesis we move $b^I$ to the left and to the
right with equal weights to get \bee\label{p11} &{}& \f{\p}{\p\mu_J}\mu^I
\Phi_\sigma (\mu )+(I\leftrightarrow J )= -\f{\nu}{2} \sum_{M}str\ig (
exp(S^\prime )\{b^J ,f^{IM}\}\mu_M \sigma\ig )\nn &{}& -\f{\nu}{2}\int
\,\sum_{L,M}(t_1 -t_2 ) str\ig (exp (t_1 S^\prime ) F^{JL}\mu_L exp (t_2
S^\prime ) f^{IM}\mu_M \sigma \ig )D^1 t + (I\leftrightarrow J ) \,. \eee The
last term on the right hand side of this expression can be shown to vanish
under the supertrace operation due to the factor of $(t_1 -t_2 )$, so that
one is left with the equation \be\label{dm1} L^{IJ}\Phi_\sigma (\mu )= -\frac
{\nu}{2} R^{IJ} (\mu )\,, \ee where \bee\label{RIJ} R^{IJ} (\mu )=\sum_{M}
str\ig ( exp(S^\prime )\{b^J ,f^{IM}\}\mu_M \sigma \ig ) +(I\leftrightarrow J
) \eee and \bee\label{LIJ} L^{IJ}=\f{\p}{\p\mu_J}\mu^I +
\f{\p}{\p\mu_I}\mu^J\,. \eee

The differential operators $L^{IJ}$ satisfy the standard $sp(2E(\sigma))$
commutation relations \be\label{lcom} [L^{IJ},L^{KL}]= - \left( {\cal
C}^{IK}L^{JL}+ {\cal C}^{IL}L^{JK}+ {\cal C}^{JK}L^{IL}+ {\cal C}^{JL}L^{IK}
\right) \,. \ee We show by induction in Appendix \ref{appc} that this algebra
is consistent with the right-hand side of the basic relation (\ref{dm1}) \ie
that \be\label{c3} [L^{IJ},\,R^{KL}]- [L^{KL},\,R^{IJ}]= -\left( {\cal
C}^{IK}R^{JL}+ {\cal C}^{JL}R^{IK}+ {\cal C}^{JK}R^{IL}+ {\cal C}^{IL}R^{JK}
\right) \,. \ee

Generally, these consistency conditions guarantee that the equations
(\ref{dm1}) express $\Phi_\sigma (\mu ) $ in terms of $R^{IJ}$ in the
following way \bee \label{ex2} \Phi_\sigma(\mu)&=& \Phi_\sigma(0)+\f
{\nu}{8E(\sigma)}\sum_{I,J=1}^{2E(\sigma)} \int_0^1 \frac{dt}{t}
(1-t^{2E(\sigma )}) (L_{IJ} R^{IJ})(t\mu ) \,, \eee provided that \be
\label{0} R^{IJ}(0)=0\,. \ee The latter condition must hold for the
consistency of (\ref{dm1}) since its left hand side vanishes at $\mu_I =0$.
In the formula (\ref{ex2}) it guarantees that the integral on $t$ converges.
In the case under consideration the property (\ref{0}) is indeed true as a
consequence of the definition (\ref{RIJ}).

Taking into account {\it Lemma 5} and the explicit form of $R^{IJ}$
(\ref{RIJ}) one concludes that the equation (\ref{ex2}) expresses uniquely
the supertrace of special polynomials via the supertraces of polynomials of
lower degrees or via the supertraces of special polynomials of the same
degree with a lower number of cycles of even length provided that the $\mu$
independent term $\Phi_\sigma(0)$ is an arbitrary solution of {\it GLC}.
This completes the proof of {\it Theorem 2}.

\noindent {\it {\bf Comment 1:} The formulae (\ref{ex1}) and (\ref{ex2}) can
be effectively used in practical calculations of supertraces of particular
elements of $SH_N (\nu)$.}

\noindent {\it {\bf Comment 2:} Any supertrace on $SH_N (\nu )$ is determined
unambiguously in terms of its values on the group algebra of $S_N$.}

\noindent {\it Corollary:} Any supertrace on \A is $\rho$-invariant,
$str(\rho(x))=str(x)$ $\forall x \in SH_N(\nu)$, for the antiautomorphism
$\rho$ (\ref{ant}).

\noindent This is true due to the {\it Comment 2} because $\sigma$ and
$\sigma^{-1}=\rho (\sigma )$ belong to the same conjugacy class of $S_N$ so
that $str(\rho (\sigma ))=str(\sigma)$.

\section{Conclusions.}\label{sec6}

In this paper we have shown that the algebras $SH_N (\nu )$ can be endowed
with $q(N)$ independent supertrace operations where $q(N)$ is the number of
partitions of $N$ into a sum of odd positive integers. We hope to apply the
supertraces constructed in this paper to the analysis of the invariant forms
of $SH_N(\nu)$. Although a definition of the supertraces on \A behaves
regularly with the parameter $\nu$ (in particular, the number of supertraces
$q(N)$ is $\nu$ independent) one can expect that this is not the case for the
related bilinear forms which can degenerate for some special values of $\nu$
thus giving rise to ideals of \A as it happens \cite{14} for the simplest
case of $N=2$. The analysis of the structure of these ideals is a challenging
problem important for various application of $SH_N(\nu)$, including analysis
of its representations. We are going to study this problem for some lower
values of $N>2$ in the future publication.

In conclusion let us note that the method of the analysis of supertraces
presented in this paper is rather general. Practically, the only information
of the specific structure of $SH_N(\nu)$ is that {\it Lemma 5} is true.
Hopefully one can use the analogous methods for the analysis of supertraces
of other associative algebras.

\vskip 5 mm \noindent {\bf Acknowledgements} \vskip 3 mm \noindent Authors
are very grateful to M.~Soloviev for useful discussions. The research
described in this publication was made possible in part
by Grant $\mbox{N}^{\mbox{\underline o}}$
MQM300 from the International Science
Foundation and Government of Russian Federation.
This work was supported in part by
the Russian Basic Research Foundation, grant 93-02-15541, and INTAS grant
93-0633.

\appendix

\vskip 8 mm \noindent {\large \bf APPENDICES} \vskip 3 mm

\setcounter{equation}{0} \renewcommand{\theequation}{A\arabic{equation}}

\section{Independence $G_N$ of $\nu$.}\label{appa} For the case $\nu =0$ it
was argued in section \ref{sec4} that (\ref{mm}) possesses $q(N)$ independent
solutions. Let us now consider the case $\nu \neq 0$. By induction on a
number of cycles of even length $e=E(\sigma)$ we show that given $\sigma$
with $E(\sigma)=e \geq 1$ there is only one independent equation on
$str(\sigma)$ provided that all equations (\ref{mm}) with $E(\sigma)=e^\prime
<e$ are resolved. In this proof we set $\nu =1$ that does not lead to the
loss of generality due to the scaling property (\ref{scal}). The first step
of the induction consists of the observation that there are no equations for
the case $E(\sigma )=0$.

Let us consider the case where there are two equations (\ref{eqss}) on
$str(\sigma)$ for some $\sigma$. This is only possible if $\sigma= c_1 c_2
\sigma^\prime$ where $c_1$ and $c_2$ are some cycles in the decomposition of
$\sigma$ such that $|c_1|=2k$, $|c_2|=2l$ and $k \neq l$. Note that
$E(\sigma^\prime)=E(\sigma)-2=e-2$.

Without loss of generality let us set \bee \label{rclrcl} c_1= K_{12}
K_{23}\,...\,K_{(2k-1)\,2k}\,,\qquad c_2=K_{(2k+1)\,(2k+2)}
\,...\,K_{(2k+2l-1)\,(2k+2l)}\,, \eee \be b^\alpha_1= \frac 1 {\sqrt {2k}}(
a^\alpha_{1}-a^\alpha_{2}+...-a^\alpha_{2k})\,,\qquad b^\alpha_2= \frac 1
{\sqrt {2l}}(a^\alpha_{2k+1}-a^\alpha_{2k+1}+...-a^\alpha_{2k+2l})\,. \ee
 Also we introduce \bee\label{c} c&=&K_{1\, (2k+1)}c_1 c_2= K_{12}
K_{23}\,...\,K_{(2k+2l-1)\,(2k+2l)} \eee and \bee b^\alpha &=&\frac 1 {\sqrt
{2k+2l}} ( \sqrt {2k} b^\alpha_1+ \sqrt {2l} b^\alpha_2) = \frac 1 {\sqrt
{2k+2l}} (a^\alpha_{1}-a^\alpha_{2}+\ldots -a^\alpha_{2k+2l}). \eee

The corresponding equations (\ref{eqss}) take the form \bee
str(\sigma)=-str\ig(\big( [b^0_1,b^1_1]-1\big)\sigma\ig) \label{1} \eee and
\bee str(\sigma)=-str\ig(\bigg( [b^0_2,b^1_2]-1\bigg)\sigma\ig). \label{2}
\eee Using the following simple identity which holds for any trace on $S_N$,
$$ str\bigg(\frac 1{2k}\sum_{p=1}^{2k}\sum_{q=2k+1}^{2k+2l}K_{p\,q}\sigma
\bigg)= 2l\,str(K_{1\,(2k+1)}\sigma) $$ one can rewrite the right hand side
of (\ref{1}) as \be\label{1c} str\bigg(\ig ([b^0_1,b^1_1]-1\ig )\sigma
\bigg)= str\bigg (\ig ([b^0_1,b^1_1]-1- \frac
1{2k}\sum_{p=1}^{2k}\sum_{q=2k+1}^{2k+2l}K_{p\,q}\ig)\sigma\bigg )+
2l\,str(K_{1\,(2k+1)}\sigma). \ee

The direct analysis based on the commutation relations (\ref{gcom}) and
(\ref{A}) then shows that the first term on the right hand side of (\ref{1c})
is the supertrace of a linear combination of permutations which all contain
the cycle $c_2$ in their decompositions. The second term is the supertrace of
the permutation which contains the cycle $c$ (\ref{c}) in its decomposition.
It is easy to see that for each of these terms the number of cycles of even
length is $E(\sigma )-1$. This allows us to apply the equation (\ref{eqss})
to each of this terms due to the inductive hypothesis. We identify $c_0$ with
$c_2$ and $c$, respectively, for the first and second terms on the right hand
side of (\ref{1c}). As a result the equation (\ref{1}) turns out to be
transformed to the form \bee\label{ij} str(\sigma) &=& str\bigg(\ig
([b^0_2,b^1_2]-1\ig ) \ig ([b^0_1,b^1_1]-1- \frac
1{2k}\sum_{p=1}^{2k}\sum_{q=2k+1}^{2k+2l}K_{p\,q}\ig ) \sigma\bigg)\nn &+&
str\bigg( ([b^0,b^1]-1) 2lK_{1\,(2k+1)}\sigma\bigg)\,, \eee Analogously one
obtains for (\ref{2}) \bee\label{ji} str(\sigma) &=& str\bigg(\ig
([b^0_1,b^1_1]-1\ig ) \ig ([b^0_2,b^1_2]-1- \frac
1{2l}\sum_{p=1}^{2k}\sum_{q=2k+1}^{2k+2l}K_{p\,q}\ig ) \sigma\bigg)\nn &+&
str\left( ([b^0,b^1]-1) 2kK_{1\,(2k+1)}\sigma\right). \eee

Let us prove that the difference of the right hand sides of (\ref{ij}) and
(\ref{ji}) vanishes. With the aid of the simple consequence of the $S_N$
invariance $$ \frac 1{4kl}str\bigg (
[b^0_i,b^1_i]\sum_{p=1}^{2k}\sum_{q=2k+1}^{2k+2l}K_{p\,q}\sigma\bigg )=
str\ig ([b^0_i,b^1_i]K_{1\,(2k+1)} \sigma\ig ), \qquad i=1,2 $$ this
difference can be transformed to the form \bee\label{dif} {X}=
str\left(\left(2k\,[b^0_1,b^1_1]-2l\,[b^0_2,b^1_2]\right)K_{1\,(2k+1)}
\sigma\right)\,, \eee where we have taken into account that \be \label{9}
str\left([b^0,b^1] K_{1\,(2k+1)}\sigma\right)=0 \ee as a consequence of the
inductive hypothesis and {\it GLC} (\ref{ccc}) and that \be\label{90}
str\left(\left[\ig ([b^0_2,b^1_2]-1\ig ) ,\,\ig ([b^0_1,b^1_1]-1\ig
)\right]\sigma\right)=0 \ee since each term in the commutator belongs to the
group algebra of $S_N$ and commutes with $\sigma$ so that (\ref{90}) vanishes
for any supertrace on the group algebra of $S_N$.

Using the relation $b_1^\alpha=1/\sqrt{2k}\,\ig (\sqrt{2k+2l} \,b^\alpha -
\sqrt{2l}\,b_2^\alpha\ig )$ one transforms $X$ to the form \bee \label{last}
X= 2str\igg (\ig ( (k+l)\,[b^0,b^1]- \sqrt{l(k+l)}\,[b^0,b^1_2]-
\sqrt{l(k+l)}\,[b^0_2,b^1]\ig ) K_{1\,(2k+1)}\sigma\igg ). \eee Due to the
$S_N$ invariance the second term on the right hand side of (\ref{last}) can
be rewritten as \bee &{}& -2\sqrt{l(k+l)}\,\,str\bigg (\big [b^0,b^1_2\big ]
K_{1\,(2k+1)}\sigma\bigg )= -\frac {\sqrt{l(k+l)}} {k+l} \sum_{p=1}^{2k+2l}
str\bigg (c^p\left[b^0,\, b^1_2 \right]c^{-p} K_{1\,(2k+1)}\sigma\bigg )\nn
&=& -\frac {\sqrt{l}}{\sqrt {k+l}}\, str\bigg (\ig
[b^0,\,\sum_{p=1}^{2k+2l}{(-1)^p}c^pb^1_2c^{-p}\ig ] K_{1\,(2k+1)}\sigma\bigg
)= -2l\,str([b^0,b^1] K_{1\,(2k+1)}\sigma)\,. \n \eee Analogously one can
transform the third term on the right hand side of (\ref{last}). Using again
(\ref{9}) one concludes that $X=0$.

Thus it is shown that the number of solutions of (\ref{mm}) is equal to the
number of the conjugacy classes in $S_N$ with $E(\sigma)=0$. This completes
the proof of {\it Theorem 1}.

\setcounter{equation}{0} \renewcommand{\theequation}{B\arabic{equation}}

\section{Consistency for $\lambda\neq -1$}\label{appb}

Let us prove by induction that the equations (\ref{c1}) are true for any two
$\mu_1 \equiv \mu_{K_1}$ and $\mu_2 \equiv \mu_{K_2}$ such that both
$\lambda_1\equiv \lambda_{K_1}\neq -1$ and $\lambda_2\equiv \lambda_{K_2}\neq
-1$. To implement induction one selects from (\ref{start}) a part of order
$k$ in $\mu$ and observes that it contains the anticommutator of $b^L$ with a
degree $k$ polynomial in $b^M$ while the part on the right hand side of the
differential version (\ref{nm1}) of (\ref{start}) which is of the same order
in $\mu$ has the order $k-1$ as the polynomial of $b^M$. This happens because
of the presence of the commutator $[b^L ,S]$ which is a degree zero
polynomial due to the basic relations (\ref{gcom}), (\ref{A}). As a result,
the inductive hypothesis allows one to use the properties of the supertrace
provided that the above commutator is always handled as the right hand side
of (\ref{gcom}) (\ie it is not allowed to represent it again as a difference
of the second-order polynomials).

Direct differentiation with the aid of (\ref{d}) gives \bee \label{p1}
(1+\lambda_2 )\f{\p}{\p\mu_2 } \int \,(\lambda_1 t_1 -t_2 ) str\ig ( e^{t_1
S} [b^1,S]\,e^{t_2 S}\sigma\ig ) D^1 t - \ig (1 \leftrightarrow 2 \ig )
&=&\nn \left(\int \, (1+\lambda_2 )(\lambda_1 t_1 -t_2 ) str\left( e^{t_1 S}
[b^1 ,b^2 ]\,e^{t_2 S} \sigma \right )D^1 t \, - \ig (1 \leftrightarrow 2 \ig
) \right) &+&\nn \igg ( \int (1+\lambda_2 ) (\lambda_1 (t_1 +t_2 )-t_3 )
 str\ig ( e^{t_1 S} b^2 e^{t_2 S} [b^1 ,S]\,e^{t_3 S} \ig ) D^2 t\,\, - \ig
 (1 \leftrightarrow 2 \ig ) \igg ) &+& \nn \igg ( \int (1+\lambda_2 )
(\lambda_1 t_1 -t_2 -t_3 ) str \ig( e^{t_1 S} [b^1 ,S]\,e^{t_2 S} b^2 e^{t_3
 S}\sigma \ig )D^2 t \, - \ig (1 \leftrightarrow 2 \ig ) \igg)\,.&{}& \eee

We have to show that the right hand side of (\ref{p1}) vanishes. Let us first
transform the second and the third terms on the right-hand side of
(\ref{p1}). The idea is to move the operators $b^2$ through the exponentials
towards the commutator $[b^1 ,S]$ so that to use then Jacobi identities for
the double commutators. This can be done in two different ways inside the
supertrace so that one has to fix appropriate weight factors for each of
these processes. The correct weights turn out to be \bee\label{p2} D^2
t(\lambda_1 (t_1 +t_2 )-t_3 )b^2 \equiv D^2 t(\lambda_1 -t_3 (1+ \lambda_1
))b^2= \nn D^2 t\left (\igg (\f{\lambda_1 \lambda_2}{1+\lambda_2}-t_3 (1+
\lambda_1 )\igg ) \overrightarrow {b^2} + \f{\lambda_1 }{1+\lambda_2}
\overleftarrow {b^2} \right ) \n \eee and \bee\label{p3} D^2 t(\lambda_1 t_1
-t_2 -t_3 )b^2 \equiv D^2 t ((\lambda_1 +1) t_1 -1)b^2= \nn D^2 t\left ( \igg
(t_1 (1+ \lambda_1 ) -\f{1}{1+\lambda_2} \igg ) \overleftarrow {b^2} -
\f{\lambda_2 }{1+\lambda_2} \overrightarrow {b^2} \right ) \n \eee in the
second and third terms on the right hand side of (\ref{p1}), respectively.
Here the notations ${\overrightarrow A}$ and ${\overleftarrow A}$ imply that
the operator $A$ has to be moved from its position to the right and to the
left, respectively. Using (\ref{r3}) along with the simple formula
\be\label{p4} \int \, \phi (t_3 ,\ldots t_{n+1} )D^n t= \int \, t_1\phi (t_2
,\ldots t_n )D^{n-1} t \n \ee one finds that all terms which involve both
$[b^1 ,S]$ and $[b^2,S]$ cancel pairwise after antisymmetrization
$1\leftrightarrow 2$.

As a result, one is left with some terms involving double commutators which
by virtue of Jacobi identities and antisymmetrization all reduce to
\be\label{1p5} \int\,\ig (\lambda_1\lambda_2 t_1+t_2-t_1 t_2
(1+\lambda_1)(1+\lambda_2)\ig ) str\ig (exp (t_1 S) [S,[b^1 ,b^2 ]]exp (t_2
S) \sigma\ig ) D^1 t\,. \n \ee Finally one observes that this expression can
be equivalently rewritten in the form \be\label{p5} \int \, \ig
(\lambda_1\lambda_2 t_1+t_2-t_1 t_2 (1+\lambda_1)(1+\lambda_2)\ig )
\left(\f{\p}{\p t_1}-\f{\p}{\p t_2} \right) str\ig (exp (t_1 S) [b^1 ,b^2
]exp (t_2 S) \sigma\ig ) D^1 t \ee and after integration by parts cancel the
first term on the right-hand side of (\ref{p1}). Thus it is shown that the
equations (\ref{nm1}) are mutually compatible for the case $\lambda_{1,2}\neq
-1 $.

Analogously one can show that the equations (\ref{nm1}) are consistent with
(\ref{m1}). Actually, let $\lambda_1 =-1$, $\lambda_2 \neq -1$. Let us prove
that \be\label{p6} \f{\p}{\p\mu_2}str\ig ([b^1 ,exp(S)]\sigma \ig )=0 \ee
provided that the supertrace is well defined for the lower order polynomials.
The explicit differentiation gives \bee\label{p7} \f{\p}{\p\mu_2}str\ig ([b^1
,exp(S)]\sigma \ig )&\n&= \int \,str\ig ([b^1 ,exp(t_1 S)b^2 exp(t_2
S)]\sigma \ig )D^1 t \nn &\n&= (1+\lambda_2 )^{-1} str\ig ([b^1 ,(b^2 exp(S)
+ \lambda_2 exp( S)b^2 )]\sigma \ig ) +\ldots \eee where dots denote some
terms of the form $str\ig( [b^1 , B]\sigma\ig )$ involving further
commutators inside $B$, which therefore amount to some lower order
polynomials and vanish by the inductive hypothesis. As a result, one finds
\bee\label{p8} \f{\p}{\p\mu_2}str\ig ([b^1 ,exp(S)]\sigma \ig )&\n&=
(1+\lambda_2 )^{-1} str\ig ((b^2 [b^1 ,exp(S)] + \lambda_2 [b^1 ,exp( S)]b^2
)\sigma \ig )\nn &\n&+ (1+\lambda_2 )^{-1} str\ig (([b^1 ,b^2 ]exp(S) +
\lambda_2 exp( S)[b^1 ,b^2 ])\sigma \ig )\,, \eee which expression vanishes
by the inductive hypothesis too.

\setcounter{equation}{0} \renewcommand{\theequation}{C\arabic{equation}}

\section{Consistency for $\lambda =-1$}\label{appc}

In order to prove (\ref{c3}) we use the inductive hypothesis {\bf (i)}. In
this appendix we use the convention that any upper or lower indices denoted
by the same letter are automatically symmetrized, $e.g.$ $F^{II}
\stackrel{def}{=}\f 1 2 (F^{I_1 I_2}+F^{I_2 I_1})$. Let us write the identity
\be\label{p12} 0 = \sum_{M}str\ig (\ig [ exp(S^\prime ) \{ b^I ,f^{IM}\}\mu_M
,b^Jb^J\ig ] \sigma\ig ) -(I \leftrightarrow J) \ee which holds due to {\it
Lemma 5} for all terms of degree $k-1$ in $\mu$ with $E(\sigma) \leq l+1$ and
for all lower order polynomials in $\mu$ (one can always move $f^{IJ}$ to
$\sigma$ combining them into a combination of elements of $S_N$ analyzed in
{\it Lemma 5}).

The straightforward calculation of the commutator on the right-hand-side of
(\ref{p12}) gives $0 = X_1+X_2+X_3$, where \bee\label{1p13}
X_1&=&-\sum_{M,L}\int\,str \left ( exp(t_1 S^\prime ) \{b^J , F^{JL} \} \mu_L
 exp(t_2 S^\prime ) \{b^I ,f^{IM}\}\mu_M \sigma \right) D^1 t -(I
 \leftrightarrow J)\,,\nn X_2&=&\sum_M str \left ( exp(S^\prime ) \ig
 \{\{b^J, F^{IJ}\},f^{IM} \ig\} \mu_M \sigma \right)-(I \leftrightarrow
 J)\,,\nn X_3&=& \sum_M str \left ( exp(S^\prime ) \ig
 \{b^I,\{b^J,[f^{IM},b^J]\} \ig \}\mu_M \sigma \right)-(I \leftrightarrow
 J)\,. \eee The terms bilinear in $f$ in $X_1$ cancel due to the
antisymmetrization ($I\leftrightarrow J$) and the inductive hypothesis {\bf
(i)}. As a result, one can transform $X_1$ to the form \bee\label{p13} X_1 =
\left (- \f 1 2 \left [ L^{JJ},\,R^{II}\right ] +2 str\ig ( e^{S^\prime }
\{b^I ,f^{IJ}\}\mu^J \sigma\ig ) \right) - (I \leftrightarrow J). \eee

Substituting $F^{IJ}={\cal C}^{IJ}+\nu f^{IJ}$ and $f^{IM}=\nu^{-1}
([b^I,b^M] -{\cal C}^{IM})$ one transforms $X_2$ to the form \bee\label{x2}
X_2 &=& 2{\cal C}^{IJ}R^{IJ} -2 \left( str \ig ( e^{S^\prime } \{b^J
 ,f^{IJ}\}\mu^I \sigma \ig ) -(I \leftrightarrow J) \right)+Y, \eee where
\bee\label{Y} Y= str\left ( e^{S^\prime } \ig \{ \{b^J,
 f^{IJ}\},[b^I,\,S^\prime] \ig\} \sigma \right) -(I \leftrightarrow J)\,.
 \eee Using that \bee\label{S} str \left ( exp(S^\prime ) \left[ P
 f^{IJ}Q,\,S^\prime \right] \sigma \right) =0 \eee provided that the
inductive hypothesis can be used, one transforms $Y$ to the form
\bee\label{x2x2} Y= str \bigg ( e^{S^\prime } \bigg( &-&[f^{IJ},\, (b^I
 S^\prime b^J + b^J S^\prime b^I )] {} - b^I [f^{IJ},\, S^\prime] b^J -b^J
 [f^{IJ},\, S^\prime] b^I\nn & + & [f^{IJ},\, \{b^I\,, b^J\}] S^\prime \bigg)
 \sigma \bigg). \eee

Let us rewrite $X_3$ in the form $X_3=X_3^{s}+X_3^{a}$ where \bee X_3^s= \f 1
2 \sum_M str \left ( e^{S^\prime } \ig ( \ig\{b^I,\{b^J,[f^{IM},b^J]\}\ig \}
 + \ig\{b^J,\{b^I,[f^{IM},b^J]\}\ig \} \ig ) \mu_M \sigma \right) - (I
 \leftrightarrow J)\,,\nn X_3^a= \f 1 2 \sum_M str \left ( e^{S^\prime } \ig
 ( \ig\{b^I,\{b^J,[f^{IM},b^J]\}\ig \} - \ig\{b^J,\{b^I,[f^{IM},b^J]\}\ig \}
 \ig ) \mu_M \sigma \right) - (I \leftrightarrow J)\,. \nonumber \eee With
 the aid of the Jacobi identities $[f^{IM},b^J]-[f^{JM},b^I]=[f^{IJ},b^M]$
one expresses $X_3^s$ in the form $$ X_3^s= \f 1 2 str \left ( e^{S^\prime}
\left ( \{b^I,b^J\}[f^{IJ}, S^\prime] + [f^{IJ}, S^\prime]\{b^I,b^J\} +2
 b^I[f^{IJ}, S^\prime]b^J + 2 b^J[f^{IJ}, S^\prime]b^I \right ) \sigma \right
). $$ $X_3^a$ can be transformed to the form \bee\label{100} X_3^a= \f 1 2
\sum_M str \left ( e^{S^\prime} \left[ F^{IJ},\, [f^{IM},b^J]\right ] \mu_M
 \sigma \right ) -(I \leftrightarrow J). \eee

By virtue of the substitutions $F^{IJ} ={\cal C}^{IJ}+\nu f^{IJ}$ and
$f^{IM}=\nu^{-1} ( [b^I,b^M]-{\cal C}^{IM})$ in (\ref{100}) one finds after
simple transformations that $Y+X_3=0$. {}From (\ref{p13}) and (\ref{x2}) it
follows then that the right hand side of (\ref{p12}) equals to $\f 1 2
([L^{II},\, R^{JJ}]-[L^{JJ},\, R^{II}]) + 2{\cal C}^{IJ} R^{IJ}$. This
completes the proof of the consistency conditions (\ref{c3}).

\end{document}